\documentclass[12pt]{iopart}

\usepackage{bm}
\usepackage{natbib}
\usepackage{har2nat}   
\setcitestyle{aysep={,}} 
\usepackage{color}
\usepackage{hyperref}
\usepackage{mathrsfs}
\usepackage{graphicx}
\usepackage{epstopdf}
  \expandafter\let\csname equation*\endcsname\relax
  \expandafter\let\csname endequation*\endcsname\relax
\usepackage{amstext}
\usepackage{amsmath} 
\usepackage{mathrsfs}
\usepackage{amssymb}
\usepackage[pagewise, displaymath, mathlines]{lineno}
\usepackage{color}

\begin{document}

\title[Manuscript for Meas. Sci. and Technol.]{Error Propagation Dynamics of PIV-based Pressure Field Calculations: How well does the pressure Poisson solver perform inherently?}

\author{Zhao Pan$^{1}$, Jared Whitehead$^{2}$, Scott Thomson$^{1}$ and Tadd Truscott$^{3}$}
\address{$^{1}$Department of Mechanical Engineering, Brigham Young University, UT, USA \\ 
$^{2}$Mathematics Department, Brigham Young University, UT, USA \\ 
$^{3}$Department of Mechanical and Aerospace Engineering, Utah State University, UT, USA}
\ead{panzhao0417@gmail.com \& taddtruscott@gmail.com}
\vspace{10pt}
\begin{indented}
\item[] Jan 2016
\end{indented}

\begin{abstract}
Obtaining pressure field data from particle image velocimetry (PIV) is an attractive technique in fluid dynamics due to its noninvasive nature. The application of this technique generally involves integrating the pressure gradient or solving the pressure Poisson equation using a velocity field measured with PIV. However, very little research has been done to investigate the dynamics of error propagation from PIV-based velocity measurements to the pressure field calculation. 
Rather than measure the error through experiment, we investigate the dynamics of the error propagation by examining the Poisson equation  directly. We analytically quantify the error bound in the pressure field, and are able to illustrate the mathematical roots of why and how the Poisson equation based pressure calculation propagates error from the PIV data. The results show that the error depends on the shape and type of boundary conditions, the dimensions of the flow domain, and the flow type. 

\end{abstract}

\vspace{2pc}
\noindent{\it Keywords}: PIV, pressure field calculation, error propagation, error estimation, Poisson equation, boundary conditions.

\submitto{\MST}

\section{Introduction}

Accurate pressure and velocity measurements are critical for experimental fluid dynamics. Historically, flow velocity is measured using techniques including hot wire anemometry \cite{ho1998micro}, and laser doppler velocimetry \cite{durst1981principles}. More recently, digital imaging techniques such as Particle Image Velocimetry (PIV) \cite{willert1991digital} and Particle Tracking Velocimetry (PTV) \cite{adamczyk19882} have become effective techniques that continue to be improved. The ability of techniques like PIV to non-invasively capture accurate data makes them very appealing.  A natural extension of this approach is to non-intrusively quantify the pressure field using the Navier-Stokes equation and the velocity field from PIV measurements. 

Early efforts at noninvasive pressure estimates can be traced back to \citet{schwabe1935druckermittlung} and \citet{imaichi1983numerical}. 
However, due to the technical limitations of their imaging technique (i.e., low spatial and temporal resolution, etc.) and consequently large error in the velocity field measurement, the pressure calculations were not accurate enough to ensure quantitative confidence.

After more than 20 years of development, PIV has become a standard non-invasive velocity field measurement technique \cite{adrian2005twenty}. Continual improvement has led to high temporal and spatial resolution for modern PIV techniques and even time-resolved volumetric PIV  \cite{elsinga2006tomographic, belden2010three,scarano2012tomographic}. Several groups have revisited velocity-field-based pressure calculation techniques and applied them to many different areas. \citet{de2012instantaneous} reported their work on applying high-speed PIV to planar pressure calculations in a turbulent flow field. Stereoscopic and tomographic PIV systems were used to measure out-of-plane velocity components in a flow passing over a square pillar (Re=9,500).  Van Oudheusden et al. \citeyearpar{van2007evaluation, van2008principles} extended the previous work to compressible flows. PIV data from an airfoil in a supersonic flow and shock wave boundary layer interactions were used to successfully estimate the corresponding pressure field. 

 {\color{black} More applications are outlined in Table \ref{table}. Most of these studies report the errors in the pressure field calculation by comparing the analytical, numerical and/or experimental results. Some studies (e.g., \citet{villegas2014evaluation}) provided an estimation of the error in the pressure field, based on the methods provided by \citet{ragni2009surface}, and \citet{de2012instantaneous}, which gave serious attempts to translate the uncertainties in the velocity to pressure quantitatively. However, these studies did not provide analytical insights that could be used for error estimation before experiments.}

Unfortunately uncertainties in the PIV-based velocity field measurement will always propagate to contaminate the resulting pressure field calculation. Researchers have noticed this issue and several techniques have been developed to reduce the errors in the resulting pressure field. One popular strategy is to average several pressure calculations along different integral paths by taking advantage of the scalar property of the pressure field (the integrated pressure value at an arbitrary location in the flow field is independent of the integral path). \citet{baur1999piv} directly integrated a simplified Navier-Stokes equation with an explicit scheme. They utilized time-resolved PIV data to determine the pressure of a turbulent flow passing over a wall. At each nodal point, four integrals were calculated from neighboring nodes and averaged to formulate the pressure estimation.  However, they only commented on the accuracy of the PIV, not that of the pressure estimation. 


A further reduction in the error accumulation from the uncertainties in PIV data was implemented by \citet{liu2006instantaneous}. They proposed an omni-directional integration scheme to directly integrate the pressure gradient from a virtual boundary outside the flow field. For an $M \times N$ mesh, the pressure value at each nodal point is integrated along $2(M+N)$ different paths, and the mean value of these  $2(M+N)$ integrals is used as the estimation of the local pressure. This approach leads to significant cancellations if the error is truly random. This method was validated using a synthetic flow and then applied to a cavity flow. This approach is likely the most capable of removing the most significant portion of the random error. \citet{dabiri2013algorithm} proposed an algorithm that used the median of the pressure calculated by the Poisson solver along eight paths to estimate the local pressure at each point in the field. To reduce the uncertainties in the velocity field from the PIV, a temporal filter was utilized to cancel the inherent noise, and this approach was applied to the flow around free swimmers (e.g.,  jellyfish and lamprey). Taking advantages of the scalar property of the pressure field improves the accuracy of the pressure calculation, however, these studies 
provide little insight into how the error propagates from the velocity field to the pressure field. 

In order to better understand the performance and error properties in the PIV-based pressure calculation, \citet{charonko2010assessment} reviewed and evaluated different factors (i.e., integral method, governing equations, spatial and temporal resolutions, and velocity field smoother) of calculation schemes used in the PIV-based pressure acquisition. Two unsteady synthetic flows with exact solutions and a set of PIV and pressure data from experiments were employed for benchmarking the pressure solvers with various error levels in the velocity fields. In their paper, the authors reported that the Poisson solvers are sensitive to all the aforementioned factors, but to varying levels (the resulting error can vary from less than 1\% to more than 100\%).  They also report that the error in the pressure calculation is highly dependent on the flow type, which implies that there is no optimal method for every flow type. Their study 
provides several significant contributions to the community (e.g.,  pressure solver can be very sensitive to the error in the velocity field and the boundary), but it does not provide any rigorous physical or mathematical insight into the error propagation. 

In a recent work by \citet{azijli2016posteriori}, the uncertainty propagation of the PIV-based pressure calculation is discussed in a Bayesian estimation framework. The statistical error profile of the pressure field is estimated based on certain prior knowledge of the velocity field (e.g.,  divergence free or maximum/minimum of the velocity field), and an assumption that the distribution is Gaussian. Numerical and physical experiments were conducted to validate this Bayesian method, which provide a practical solution for error quantification. However, this method requires prior information of the flow field, and does not provide insight into how the error propagates from the flow field to the pressure calculation.
 
In this paper, we first clearly specify the error-contaminated Poisson problem raised by the pressure field calculation from noisy PIV experiments. In section 2 and 3 this engineering problem is translated into an applied mathematical one, specifically by obtaining bounds on solutions of a Poisson equation. In section 4, we present rigorous bounds on the error in the pressure calculation relative to the error inherent from the PIV measurements. Several typical cases are shown as examples. In section 5, we discuss the  limitations and practical uses of this work. {\color{black} The analytical results introduced in this paper are not only error bounds that provide insight into the error propagation dynamics of the PIV-based pressure calculation, but they can also provide guidelines for experimental design. Moreover, \emph{a prior} error estimation can be potentially predicted even before experiments based on the analysis presented herein. }

\section{Problem statement}
In general, there are two types of popular schemes to calculate the pressure field: i) directly integrate the pressure gradient derived from the Navier-Stokes equation
(e.g.,  \citet{liu2006instantaneous}); ii) solve the pressure Poisson equation (e.g.,  \citet{de2012instantaneous}), which is more commonly used. Here, we focus on how the error in the velocity data propagates to the pressure field through the latter scheme. 

Rearranging the incompressible non-dimensionalized Navier-Stokes equation ({\color{black} all the variables and equations hereafter are non-dimensionalized}) and applying divergence on both sides, the pressure Poisson equation reads
\begin{equation}
\label{eq:NSeq}
\nabla^2 p = -\nabla \cdot \left(  \frac{\partial \bm{u}}{\partial t} + \left( \bm{u} \cdot \nabla \bm{u} \right)- \frac{1}{Re} \nabla^2 \bm{u} \right)=f(\bm{u}) \qquad in \ \Omega, 
\end{equation}
where $p$ is the pressure field, $\bm{u}$ denotes the velocity field, $\Omega$ is the flow domain, and $Re$ is the Reynolds number. When $Re$ is large, the viscous term can be neglected \cite{van2013piv,de2012instantaneous}.  The vector function ($f$) of the velocity field ($\bm{u}$) is called data ({\color{black}to avoid confusion, in this paper ``data'' is used as the term for the right hand side of a Poisson equation and its boundary conditions, while the experimental data (velocity vector field) from PIV is called experimental ``results'' or PIV ``results'' instead}). With certain boundary conditions, for example,
\begin{equation}
\label{eq:DBC}
p = \frac{1}{2} \left( \bm{u}^2 - \bm{u}^2_{\infty} \right) = h(\bm{u}) \qquad \ on \  \partial \Omega,
\end{equation}
and/or
\begin{equation}
\label{eq:NBC}
\nabla p \cdot \bm{n} =   - \frac{\partial \bm{u}}{\partial t} - \left( \bm{u} \cdot \nabla \bm{u} \right)+ \frac{1}{Re} \nabla^2 \bm{u} = g(\bm{u}) \qquad on \ \partial \Omega, 
\end{equation}
the pressure field can be found by solving (\ref{eq:NSeq}). Here,  $h$ and $g$ are the data on the Dirichlet boundary (typically applied to the steady irrotational region of a flow with Bernoulli's equation, especially in the far field) and Neumann boundary (commonly used on a wall boundary), respectively, which are functions of the velocity. 

In engineering practice, experiments always introduce systematic bias and/or random errors, which are usually unknown, and thus called uncertainties in the PIV community. These uncertainties will lead to a contaminated pressure calculation (denoted by $\tilde p$). The uncertainties in the pressure calculation are also unknown, which can cause even more frustration. Regardless of the physical meaning, from now on in this paper we will call uncertainties error for convenience. If we denote the error in the data of the pressure Poisson equation as $\epsilon_f$, then $\tilde p$ solves the equation with the error included 
\begin{equation}
\label{eq:PDEwithError}
\nabla^2 \tilde p= f(\bm{u})+\epsilon_f \qquad in \  \Omega. 
\end{equation}
Similarly, $\tilde{p}$ satisfies the error-contaminated boundary conditions: 
\begin{equation}
\label{eq:DBCwithError}
\tilde p = h(\bm{u}) +\epsilon_h \qquad \ on \  \partial \Omega,
\end{equation}
and/or
\begin{equation}
\label{eq:NBCwithError}
\nabla \tilde p \cdot \bm{n} = g(\bm{u}) +\epsilon_g \qquad on \ \partial \Omega, 
\end{equation}
where, $\epsilon_h$ and $\epsilon_g$ are the error on the Dirichlet and Neumann boundaries, respectively. 

Based on this problem statement, we aim to answer a question rising from engineering practice \emph{---~Question 1: How do the errors from the experimental results $\epsilon_f$, $\epsilon_h$ and/or $\epsilon_g$ affect the errors in the contaminated pressure field $\tilde{p}$?} {\color{black} Herein we address the error introduced by experiments only (e.g., random and systematic error from PIV experiments, unrealistic assumptions such as 2D modeling for 3D, quasi-steady, etc.), rather than numerical errors introduced by the Poison solver implementation (e.g., truncation error, etc.). }

\section{Modeling of the error propagation }
We now present how to translate from an engineering problem (\emph{Question 1}) to a tractable applied mathematical one (\emph{Question 3}, see below).

Let's consider $\epsilon_f$, $\epsilon_h$ and $\epsilon_g$ as perturbations to the data of the Poisson equation. Perturbing the data ($f, g,$ and/or $h$) is mathematically equivalent to propagating error from the data to the pressure field. This means that \emph{Question 1} can be rewritten as  \emph{---~Question 2: Whether and how the solution $p$ continuously depends on the data $f(\bm{u})$, $g(\bm{u})$, and/or $h(\bm{u})$? }

Assuming the uncertainty contaminated pressure field can be separated as $\tilde{p} = p + \epsilon_p$, where $\epsilon_p$ is the error in the calculated pressure field, and taking advantage of the linear property of the Laplace operator and subtracting equation (\ref{eq:NSeq}) from (\ref{eq:PDEwithError}) leads to
\begin{equation}
\label{eq:PDE_error}
\nabla^2 \epsilon_p =\epsilon_f \qquad in \ \Omega,
\end{equation}
which is a Poisson equation with respect to the error in the pressure field. Similarly, the boundary conditions read 
\begin{equation}
\label{eq:DBC_error}
\epsilon_p = \epsilon_h \qquad on \ \partial\Omega,
\end{equation}
and/or
\begin{equation}
\label{eq:NBC_error}
\nabla \epsilon_p \cdot \bm{n} = \epsilon_g \qquad on \ \partial\Omega.
\end{equation}

Since the error in the data ($\epsilon_f$, $\epsilon_h$, and $\epsilon_g$) are unknown, we do not expect to calculate the error at every specific location in the pressure field ($\epsilon_p$). However, it is possible to estimate the average \emph{error level} of the pressure field over the entire domain with equation (\ref{eq:PDE_error}), (\ref{eq:DBC_error}) and (\ref{eq:NBC_error}). To measure the level of the error, we define the $L^2$ norm in a domain, for example the error level in the pressure field as 
\begin{equation}
\label{eq:L2norm}
||\epsilon_p||_{L^2(\Omega)} = \sqrt{  \frac{ \int \epsilon_p^2 d\Omega }{|\Omega|} } ,
\end{equation}
where $|\Omega|$ is the length, area or volume of the domain, depending on the dimension of the flow field. In physical terms, the $L^2$ norm defined in equation (\ref{eq:L2norm}) measures the power of the errors per unit space, and thus we give it the term ``error level'' hereafter.  

With the defined error level, \emph{Question 2} can be transformed into \emph{---~Question~3: Whether and how $||\epsilon_p||_{L^2(\Omega)}$ is bounded by $||\epsilon_f||_{L^2(\Omega)}$, $||\epsilon_g||_{L^2 (\partial \Omega)}$, and/or $||\epsilon_h||_{L^2(\partial \Omega)}$ for the Poisson problem given by equation (\ref{eq:PDE_error}), (\ref{eq:DBC_error}), and/or (\ref{eq:NBC_error})?}

From \emph{Question 1}, to \emph{2}, and then \emph{3}, we have been able to transform a typical engineering problem to a well defined applied mathematical one: estimate the bounds of the solution to a Poisson boundary value problem (BVP) with respect to $\epsilon_p$, which is actually a measure of the error in the pressure field.

\section{Results}
\label{results}
In this section we show that the error level can be bounded in the pressure field, given the i) geometry and ii) scale of the domain, iii) type of the boundary conditions, as well as the iv) error level in the data (in the field and on the boundary) utilizing the Poincare and Cauchy-Schwartz inequalities (see  \ref{appA} for details). The results are  independent of the numerical scheme of the Poisson solver, i.e. the choice of the numerical scheme may introduce additional errors not accounted for in the present analysis. The results are general and thus work for any dimension of the domain (i.e., two-dimensional(2D) or three-dimensional(3D) flow). 


Bounds on the error for several cases with different boundary condition settings are discussed. These cases are not only typical in engineering practice but also convenient for unveiling the mathematical insights of the error propagation dynamics. Within each case study, we will validate the analytical results with numerical simulations first. Then the dynamics of the uncertainty propagation through the pressure Poisson equation will be discussed based on analysis from a flow field with more general geometry (i.e., rectangular). Finally, the physical interpretation of the mathematics and suggestions for engineering practice are addressed.

\subsection{Dirichlet boundary case}
\label{sec:DBC}
Consider a domain with pure Dirichlet boundary condition, the error in the pressure field can be bounded as
\begin{equation}
\label{eq:DBCboundGeneral}
||\epsilon_p||_{L^2(\Omega)} \leq C_D||\epsilon_f||_{L^2(\Omega)} + ||\epsilon_h||_{L^{\infty}(\partial\Omega)},
\end{equation} 
where $C_D$ is the Poincare constant, which is related to the minimum positive eigenvalue of the BVP. Specifically, in engineering practice, the value of the Poincare constant is determined by the dimension, size, and shape of the domain, as well as the type of boundary conditions (\ref{appC}).  

\begin{figure}[!ht]
\centerline{\includegraphics[width=0.4\textwidth]{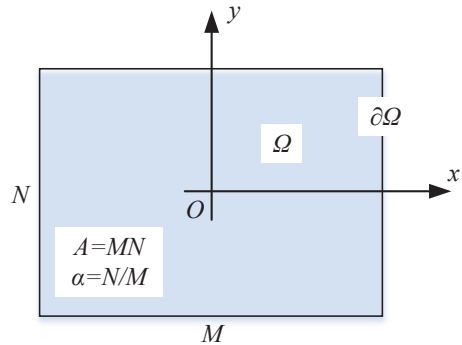}}
\caption{$N \times M$ Domain $\Omega$ with aspect ratio $\alpha$ and area $A$.}
\label{fig:NMdomain}
\end{figure}

To validate inequality (\ref{eq:DBCboundGeneral}), we consider a steady 2D potential vortex in an $ L \times L$ domain in Cartesian coordinates. The non-dimensionalized velocity field is $u = -y$, $y\in(-L/2,L/2)$; $v = x$, $x\in(-L/2,L/2)$, where $u$ and $v$ are the two components of the velocity field $\bm{u}$ in the $x$ and $y$ direction, respectively (see figure~\ref{fig:NMdomain}). Thus the pressure field is $p= (x^2+y^2)/2$. The Dirichlet boundary conditions are defined as $p = (y^2 +L^2/4) /2$, $x= \pm L/2$, and $p = (x^2 +L^2/4) /2$, $y= \pm L/2$.

We construct artificial error for the data. To test the reliability of the underlying estimate, we consider a uniformly constant error, $\epsilon_f = 2^{-4}$, and $\epsilon_h = 2^{-4}$.  The error level in the domain and on the boundary is specified identically ($||\epsilon_f||_{L^2(\Omega)} = ||\epsilon_h||_{L^{\infty}(\partial\Omega)} = 2^{-4}$). Assuming the uncertainty in the data is constant is beneficial in two ways. First, in a physical sense, constant uncertainty is associated with systematic error (i.e., $\epsilon_f = 2^{-4}$ could be equivalent to $u=-(1-2^{-6})y$ and $v=(1-2^{-6})x$ for the steady potential vortex used here). {\color{black} One of the most likely systematic errors is from} slightly inaccurate calibration in real experiments, which can introduce considerable error in the data and consequently accumulate even more error in the pressure field. Second, a constant error field will lead to a constant error level, which will make the later analyses explicit. {\color{black} We are aware that different types of error, even with the same error level (e.g., different $\epsilon_f$, yet same $||\epsilon_f||_{L^2(\Omega)}$), impact the error propagation differently, which is coupled with the profile of the velocity field. Some error in the data leads to larger error in the pressure field than others. The calibration error we choose herein yields large errors for the vortex case, making it  favorable for validation of the error bound. However, how the type of error impacts the error propagation is out of the scope of this research and will be investigated in a future study.}

We numerically solve the pressure Poisson equation with artificial error introduced \eqref{eq:PDEwithError}, using an accurate second order (five point scheme with point-wise numerical error less than $8.2\times10^{-12}$) finite difference Poisson solver ( similar to \citet{reimer2013matlab}, but with LU-decomposition as a linear system solver) . The error in the pressure field from the simulation is then compared with the analytical results inequality \eqref{eq:DBCboundGeneral}. We expect that the error from numerical simulations will be less than the prediction in \eqref{eq:DBCboundGeneral}, but generally follow similar trends. If the errors from the numerical simulations are close to the analytical prediction (i.e., slope and value), we say inequality  \eqref{eq:DBCboundGeneral} is validated and the bound is sharp. 

For the 2D square Dirichlet domain, the Poincare constant is $C_D = L^2/2\pi^2$, and inequality \eqref{eq:DBCboundGeneral} becomes
\begin{equation}
\label{eq:DBCboundSquare}
||\epsilon_p||_{L^2(\Omega)} \leq  \frac{1}{2 \pi^2}L^2 ||\epsilon_f||_{L^2(\Omega)} + ||\epsilon_h||_{L^{\infty}(\partial\Omega)}.
\end{equation} 

\begin{figure}[h!]
\begin{centering}
{\includegraphics[width=0.75\textwidth]{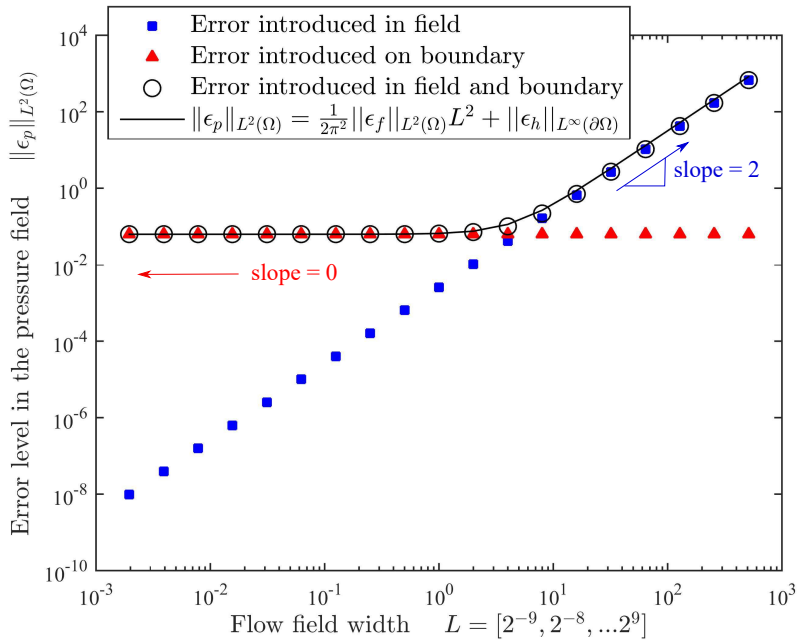}}
\caption{Error level in the pressure field versus the width of the flow field for the Dirichlet case.  The data points illustrate the error level when artificial error is introduced in the field only (blue square), on the boundary only (red triangle), and both in the field and on the boundary (black open circle). The black solid line is the bound of the error of the pressure field based on inequality (\ref{eq:DBCboundSquare}).}
\label{fig:squareD}
\end{centering}
\end{figure}

Figure \ref{fig:squareD} shows the comparison of numerical simulations and the analytical prediction. The numerical simulations are conducted based on equations \ref{eq:PDEwithError} and  \ref{eq:DBCwithError}, with the 2D potential vortex as the flow field and the constant artificial errors (introduced in the pressure Poisson equations in the field only (blue squares);  on the boundary only (red triangles); and both in the field and on the boundary (black open circles)).  Inequality (\ref{eq:DBCboundSquare}) is represented by a black solid line and is the upper bound of the error. Clearly  inequality (\ref{eq:DBCboundSquare}) fits well with the simulation results when error is introduced to both the field and boundary.


In figure \ref{fig:squareD}, when the length scale of the domain is large the uncertainty level in the pressure field is dominated by the error of the data in the field (blue squares are collapsed on black open circles), and proportional to the area of the domain ($||\epsilon_p||_{L^2(\Omega)} \sim L^2$). When the domain is small, the error in the pressure field is dominated by the error on the boundary (red triangles are collapsed on black open circles), and independent of the length scale of the domain ($||\epsilon_p||_{L^2(\Omega)} \sim L^0$).   Intuitively this makes sense as smaller domains will be more influenced by their boundaries.

When conducting PIV experiments, if the frame rate of the camera is fixed, one can customize the aspect ratio of the video, but the area of the interrogation windows is usually about the same due to best practices and the limitations of lighting and magnification (e.g., best practices of particles per pixel, particles per interrogation window, number of pixels of motion per time step, etc.). Thus, from an engineering perspective, it is important to discuss how to choose the aspect ratio when the number of pixels of the video is fixed. In order to study this physically, we vary the shape of the domain (alter the aspect ratio of only a rectangular shape due to physical restraints of the camera) to see how the error propagation dynamics is affected when the area of the domain is fixed. 

Considering a 2D $N \times M$ rectangular domain (Fig.\ref{fig:NMdomain}), inequality (\ref{eq:DBCboundGeneral}) leads to: 
\begin{equation}
\label{eq:DBCbound}
||\epsilon_p||_{L^2(\Omega)} \leq \frac{\alpha}{\pi^2(1+\alpha^2)}A||\epsilon_f||_{L^2(\Omega)} + ||\epsilon_h||_{L^{\infty}(\partial\Omega)},
\end{equation} 
where $\alpha$ is the aspect ratio ($\alpha = N/M$) of the domain, and $A$ is the area of the domain ($A=MN$). Given the error level in the data, one can use inequality (\ref{eq:DBCbound}) to estimate the error level in the pressure field. If necessary, one can also adjust the parameter settings (aspect ratio and/or area of the domain) to reduce the error propagation. 

\begin{figure}[!ht]
\centerline{\includegraphics[width=0.8\textwidth]{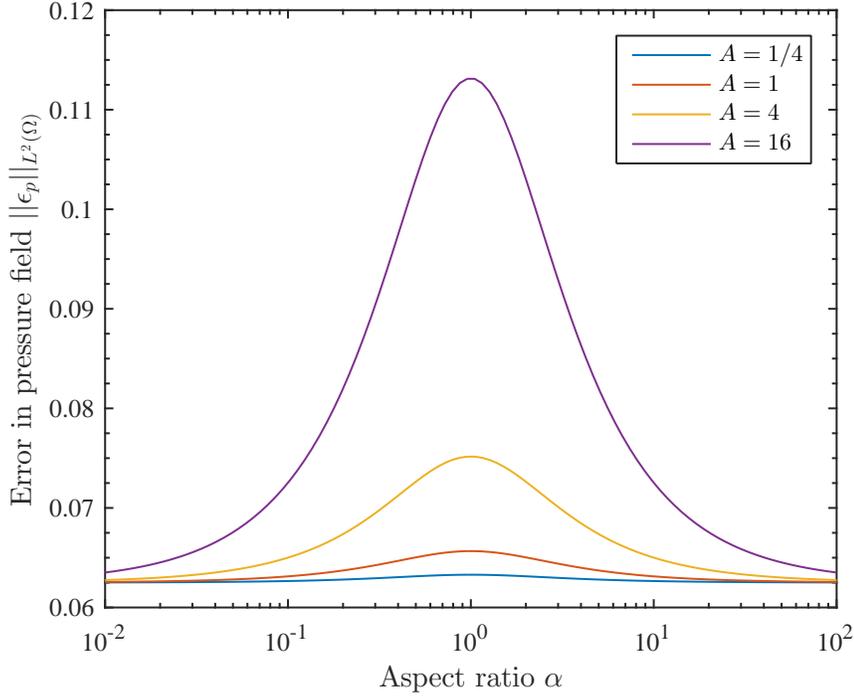}}
\caption{Error level in the pressure field versus aspect ratio of rectangular flow domains with various areas. The lines are plotted from inequality (\ref{eq:DBCbound}), with $||\epsilon_f||_{L^2(\Omega)} = ||\epsilon_h||_{L^{\infty}} = 2^{-4}$; and  $A=1/4,1,4,16$, where the lines mark the upper bound of the inequality.}
\label{fig:AspectArea}
\end{figure}

Figure \ref{fig:AspectArea} illustrates how the aspect ratio and area of the domain affect the error propagation (assuming the uncertainty level of data is $||\epsilon_f||_{L^{2}(\Omega)} = ||\epsilon_h||_{L^{\infty}(\partial\Omega)}=2^{-4}$). For each curve (fixing domain area $A$), the maximum  appears at $\alpha = 1$, which means a square PIV window is the worst case scenario if a Dirichlet boundary condition is applied. When the domain is elongated (e.g., $\alpha \rightarrow 0$) and pressure on the boundary is known, the pressure field is mainly determined by the Dirichlet boundary conditions on the longer edges, and the contribution of the error in the field and the shorter edges becomes negligible. Thus, using an elongated flow field is encouraged when precise boundary conditions are accessible, especially on the long edges. 

To compare the contributions of the uncertainty in the field and on the boundary, one can define a non-dimesional number ($R_{fb}$) which is the ratio of the coefficient of the errors in the field ($||\epsilon_f||_{L^2(\Omega)}$, inequality~(\ref{eq:DBCbound}))  and on the boundary ($||\epsilon_h||_{L^{\infty}(\partial\Omega)}$, inequality~(\ref{eq:DBCbound})). For a 2D rectangular Dirichlet domain, $R_{fb}$ reads
\begin{equation}
\label{eq:ErrorRatio}
R_{fb} = \frac{\alpha A}{\pi^2(1+\alpha^2)}. 
\end{equation} 
When $R_{fb} \ll 1$, the error on the boundary tends to dominate the error in the pressure field, with limited budget or experimental accessibility, the best way to reduce the error in the pressure field would be to improve the error on the boundary. As an example, for small areas when $A \sim 1$, then $R_{fb} \in (0,1/2\pi^2] \ll 1$, this relatively narrow interval implies that for a domain with nearly unit area, most error in the pressure field is likely contributed by the error on the boundary, while changing the aspect ratio will not affect the error in the pressure very much. However, because $R_{fb} \sim A$, the contribution of the error in the flow field increases quickly with larger domain areas.

One more comment on the Dirichlet boundary condition is that the error in the pressure field is due to the $L^\infty$ norm of the uncertainties on the boundary, which is the largest error on the boundary, rather than the average error level measured by the $L^2$ norm. It suggests that one sharp and high error peak on the boundary may significantly increase the error propagation in the pressure field. Thus, one should try to avoid outliers on the boundaries if Dirichlet boundary conditions are applied.

\subsection{Neumann case}
For a domain with Neumann boundary conditions, we can obtain the error in the pressure field using similar analyses to section~\ref{sec:DBC}. Here, we assume a zero mean error of the data in the field ($\int \epsilon_g d\Omega= 0$; see \ref{appB} for more details, {\color{black}and at the end of this section where the validity of this hypothesis is discussed}), which is the compatibility condition of the Poisson equation with pure Neumann boundary conditions. The error in the pressure field can then be bounded as 
\begin{equation}
\label{eq:NBCBoundGeneral}
||\epsilon_p||_{L^2(\Omega)} \leq C_N 
||\epsilon_f||_{L^2(\Omega)} +\sqrt{C_N C_{NB}}
||\epsilon_g||_{L^2(\partial \Omega)},
\end{equation}
where $C_N$ and $C_{NB}$ are the Poincare constants for the Neumann domain and the Neumann boundary, respectively. 

We now validate the bound introduced by inequality \eqref{eq:NBCBoundGeneral}, similar to section \ref{sec:DBC}, by considering a steady 2D potential vortex in an $L \times L$ domain. Inequality \eqref{eq:NBCBoundGeneral} becomes:
\begin{equation}
\label{eq:NBCBoundSquare}
||\epsilon_p||_{L^2(\Omega)} \leq \frac{1}{\pi^2} 
||\epsilon_f||_{L^2(\Omega)} L^2 + \frac{4}{\pi^{3/2}} ||\epsilon_g||_{L^2(\partial \Omega)}L. 
\end{equation}

We construct the same flow as in Dirichlet case, of which the non-dimensionalized velocity field is $u = -y, y\in(-L/2,L/2)$; $v = x, x\in(-L/2,L/2)$, where $u$ and $v$ are the two components of the velocity field $\bm{u}$ in the 2D Cartesian system. Thus, $f(\bm{u})=-2$, and the pressure field is $p= (x^2+y^2)/2$. To satisfy the compatibility condition of the Neumann boundary Poisson equation, the Neumann boundary conditions are $\nabla p \cdot \bm{n}=-1,x=y=-L/2$, and $\nabla p \cdot \bm{n}=1,x=y=L/2$.

Similar to section \ref{sec:DBC}, a constant artificial error is constructed: $\epsilon_f = 2^{-4}$, and $\epsilon_g = 2^{-4}$, the error level in the domain and on the boundary are constants ($||\epsilon_f||_{L^2(\Omega)} = ||\epsilon_g||_{L^2(\Omega)} = 2^{-4}$). 

\begin{figure}[!ht]
\centerline{\includegraphics[width=0.8\textwidth]{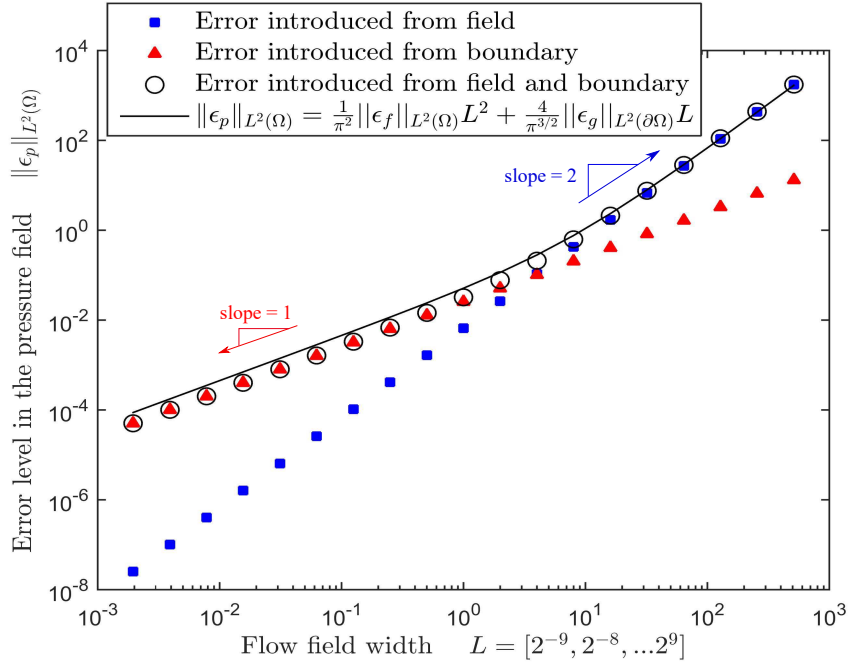}}
\caption{Error level in the pressure field versus the length scale of the flow field for the Neumann case. The data points illustrate the error level when artificial error is introduced in the field only (blue square), on the boundary only (red triangle), and in both field and on boundary (black open circle), respectively. The black solid line presents the bound of the error of the pressure field.} 
\label{fig:squareN}
\end{figure}

Introducing the error to the field only, on the boundary only, and both in the field and on the boundary, simulations agree with the theoretical analyses  (figure \ref{fig:squareN}). Error in the pressure field scales as the square of the domain length ($\sim L^2$) for large scale flow fields; however, for smaller flow fields, error scales by the domain length ($\sim L$). {\color{black} Comparing figures \ref{fig:squareD} and \ref{fig:squareN} illustrates the different trends of the error bounds due to different boundary condition settings. In addition, the amplitude of the error bounds in the pressure are significantly different even though the error level of the data are identical for each domain with varying length scale. A more detailed comparison can be found in the discussion section and figure \ref{fig:example}. }

\begin{figure}[!ht]
\centerline{\includegraphics{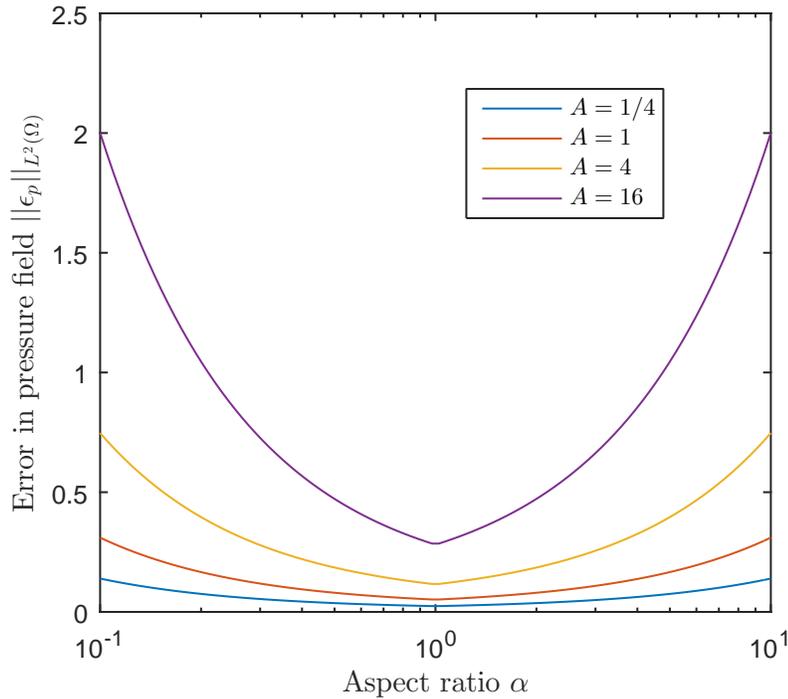}}
\caption{Error level in the pressure field versus aspect ratio of rectangular flow domains with various areas. The lines are plotted from equation \ref{eq:DBCbound}, with $||\epsilon_f||_{L^2(\Omega)} = ||\epsilon_h||_{L^{\infty}} = 2^{-4}$; and  $A=1/4,1,4,16$.}
\label{fig:AspectArea2}
\end{figure}

We also consider the more general case of a rectangle, 2D $N \times M$ field with area $A$, and aspect ratio $\alpha$. Inequality (\ref{eq:NBCBoundGeneral}) then becomes 
\begin{equation}
\label{eq:NBCbound}
||\epsilon||_{L^2(\Omega)} \leq \frac{1}{\pi^2}A\alpha^{\text{sgn}(\alpha - 1)}||\epsilon_f||_{L^2(\Omega)} + \frac{2}{\pi^{3/2}}\sqrt{A}(\alpha^{\text{sgn}(\alpha - 1)}+1)||\epsilon_g||_{L^{2}(\partial\Omega)},
\end{equation}
Figure \ref{fig:AspectArea2} shows an illustration of the error bound in the pressure field when the Neumann boundary conditions are applied. For a domain with fixed area, the square domain with $\alpha = 1$ leads to minimum error propagation. However, when an elongated domain is used, the error in the pressure may not be bounded when $\alpha \rightarrow 0 \text{\ or\ } \infty$, because the error in the pressure field is dominated by the error on the longer boundaries. Thus, in engineering practice, a square domain is recommended for Neumann boundary conditions. If an elongated domain must be used, precise Neumann boundaries should be applied to the longer boundaries, or a smaller domain should be used to reduce the error accumulation. 

Similar to section~\ref{sec:DBC} we compare the coefficients of inequality (\ref{eq:NBCbound}) and formulate the contribution ratio as
\begin{equation}
\label{eq:RatioN}
R_{fb} = \frac{\sqrt{\pi}}{2} \frac{\alpha^{\text{sgn}(\alpha-1)}}{\alpha^{\text{sgn}(\alpha-1)}+1} \sqrt{A}.
\end{equation} 
A fixed domain area, for example $A \sim 1$, yields a relatively wide interval compared with the Dirichlet case ($R_{fb} \in [\sqrt{\pi}/4,1)$). The implication is that the aspect ratio can be easily used to control the contribution from the field and boundary, depending on the specific practices of the experiments. On the other hand, $R_{fb} \sim \sqrt{A}$, meaning that the contribution ratio is proportional to the length scale of the domain, and thus not as sensitive as Dirichlet boundary conditions to the scale of the domain. 

The last comment on the pure Neumann boundary case is about the derivation of the error bound in the pressure field, and more details can be found in the appendix. The inequality \eqref{eq:NBCBoundGeneral} is obtained based on the weak or unrealistic assumption (i.e., the mean value of error in the data field is zero). Systematic error in the experiments is not necessarily a mean zero field (e.g., Gaussian errors). This could conflict with the compatibility condition and eventually render the Poisson solver intractable. Once the compatibility condition is not satisfied by the data, the solution to the Poisson equation does not even exist. One can usually get some results (we would rather not call them solutions) from a numerical Poisson solver even if the compatibility condition is not satisfied, however, the results highly depend on the numerical scheme, resolution, and convergence criteria of the numerical solver. Thus, pure Neumann boundary conditions should be avoided if possible, unless the PIV experiments have reasonably high accuracy, or the engineering application allows   Neumann boundaries only.  This tricky, but important message is brought up by very few in the literature (e.g., \citet{neeteson2015pressure} mention this but don't explain why it happens). This may be the reason why most researchers use Dirichlet or mixed boundaries; although, technically, if one can use Dirichlet BCs, Neumann BCs are also an option. We did an exhaustive literature review for the related papers published in major journals and conferences in the last five years, and found that by default the community by in large utilized Dirichlet BCs whenever possible and shied away from Neumann BCs (see table \ref{table}). {\color{black} Two studies used pure Neumann BCs, however, they either have no accessibility to Dirichlet boundaries (e.g.,  internal flows of \citet{dopplerpressure}), or a relatively small domain is used for an external flow without a confident far field assumption (e.g., \cite{villegas2014evaluation}.} {\color{black} The above statement is based on an assumption that the error level in the Neumann and Dirichlet boundary are comparable and small. However, unlike Neumann boundary conditions, which can always be imposed using PIV results, Dirichlet boundary conditions can typically be imposed only in irrotational regions using the Bernoulli equation. Any improper application of the Bernoulli equation may lead to highly uncertain (or even erroneous) Dirichlet boundary conditions and make the application of the Dirichlet boundaries unfavorable. This is indeed a dilemma that requires researchers to pay close attention to when designing an experiment; requiring that accurate boundary conditions (no matter what type) are imposed.}

\begin{table}
\caption{\label{table} Types of boundary conditions used in recent studies}
\footnotesize
\begin{tabular}{@{}ll}
\br
Type of BCs & Papers$^a$  \\
\mr
Dirichlet BCs & \citet{neeteson2015pressure}$^b$ \\
\\
Neumann BCs & \citet{dopplerpressure};\citet{villegas2014evaluation} et al.,\\
\\
Mixed BCs & \begin{tabular}[c]{@{}l@{}}\citet{jalalisendi2014particle, oren2014intraglottal, lignarolo2014experimental};  \\ \citet{de2013pressure, novara2013particle}; \\ \citet{probsting2013estimation, nila2013piv,  albrecht2013deriving}; \\ \citet{ghaemi2013turbulent, ghaemi2012piv,koschatzky2011study}; \\ \citet{moore2011two, violato2011lagrangian}, et al. and many more $^c$ \end{tabular}\\
\br
\end{tabular}\\
$^a$We only count the papers that utilize the pressure Poisson approach.\\
$^b$This paper tested both Dirichlet and Neumann boundaries for comparison.\\ 
$^c$We apologize that we cannot list the many more informative papers that used mixed boundary conditions, but we cannot have all of them listed in this table. 
\end{table}
\normalsize

\subsection{Mixed boundary conditions}
We see complicated and distinctly different error propagation dynamics simply from the boundary conditions even for these simple 2D domains. However, in engineering practice, mixed boundary conditions are more common due to the limitations and/or applications of the experiments (table~\ref{table}). We now focus on the coupled dynamics of how the geometry and boundary conditions impact the error propagation in more complicated situations (e.g., a rectangular domain with two Dirichlet boundaries, and two Neumann boundary conditions on the opposite edges of the domain, respectively).

\begin{figure}[!ht]
\centerline{\includegraphics{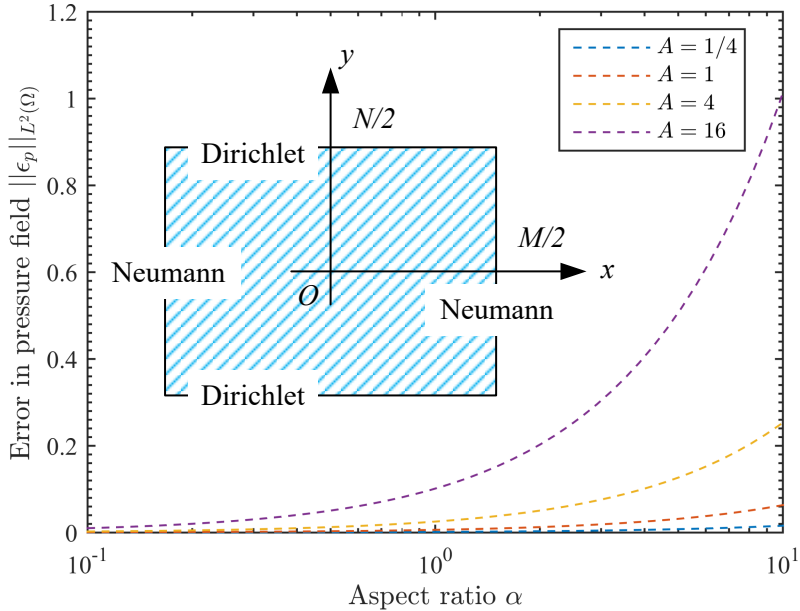}}
\caption{Error level in the pressure field versus aspect ratio of rectangular flow domains with various areas. The lines are plotted from inequality (\ref{eq:MixBCbound}), with $||\epsilon_f||_{L^2(\Omega)} =  2^{-4}$; and  $A=1/4,1,4,16$. The inset shows the boundary condition settings for the $N \times M$ domain. }
\label{fig:AspectArea3}
\end{figure}

Consider a flow in a 2D rectangular domain ($N \times M$), with mixed boundary condition ($p=h$, $y=\pm N/2 $; and $\nabla p \cdot \bm{n} = g $, $x=\pm M/2 $). This physically means that the aspect ratio can be viewed as the relative amount of the boundary dictated by a Neumann condition to that given by a Dirichlet condition.   The mixed boundary condition case can be decomposed into three parts, one that incorporates the error in the bulk of the domain, one for the error on the Neumann part of the boundary, and a third that accounts for the error on the Dirichelt part of the boundary. The analysis of the the error from the boundary terms is inherently difficult to estimate. However, for a sufficiently large convex domain we would expect the error in the interior of the domain to dominate the boundary error, and the analysis of the contribution of the error from the field is already enough to lend itself rich physical insight. We again estimate the error in the pressure field using our previous analysis: 
\begin{equation}
\label{eq:MixBCbound}
||\epsilon||_{L^2(\Omega)} \lesssim \frac{A\alpha}{\pi^2}||\epsilon_f||_{L^2(\Omega)}.
\end{equation} 
The inequality is plotted in figure~\ref{fig:AspectArea3} and shows that for a domain with constant area a larger aspect ratio (i.e., more influence from Neumann boundaries) results in large error propagation with a trend that is as fast as the pure Neumann case. 
This implies that if a mixed boundary condition is utilized in a rectangular domain, Dirichlet conditions should be used on the longer sides of the boundary to mitigate the error propagation. 

\section{Discussions}
We have shown that the upper bound of the error in the pressure field is related to the type of boundary conditions, geometry and the scale of the flow domain. The results include the explicit dependence on the geometry  (the shape and boundary of the domain is incorporated in the Poincare constant), dimension (2D or 3D), and numerical scheme of the Poisson solver. One can use these results to design and minimize the error in an experiment before it is performed. For example, one can adjust the aspect ratio, area of the domain, and the type of boundary conditions to reduce the error propagation from the velocity field to the pressure field based on the reasoning outlined in section~\ref{results}.

\begin{figure}[!th]
\centerline{\includegraphics{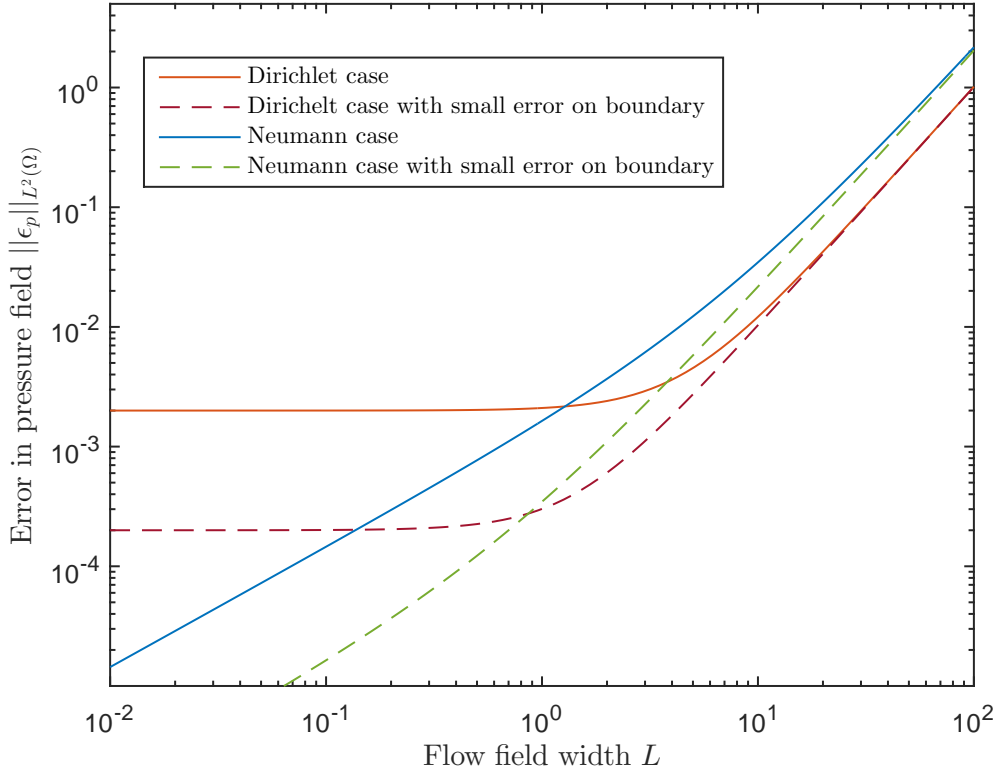}}
\caption{Bounds of the $L^2$-norm of error in a pressure field. The red and blue curves indicate the highest possible uncertainty level in the pressure field with the same level of error introduced in the data ($2^{-3}$ on boundary and in field) for Dirichlet and Neumann cases, respectively. The  purple and green dashed lines show Dirichlet and Neumann case with the same uncertainty level as other cases in the flow field, but with less error ($2^{-4}$) on the boundaries.}
\label{fig:example}
\end{figure}

We can illustrate how one might use this information by using a simple example to present how to choose boundary conditions.  Assume a square domain is used in a PIV experiment and the error level in the data and on the boundary is given and plotted in figure~\ref{fig:example}. We introduce the same error to the data in the field and the boundary for the pure Dirichlet and pure Neumann cases ($||\epsilon_f||_{L^2(\Omega)} = 2^{-3}$, $||\epsilon_h||_{L^2(\partial\Omega)} = ||\epsilon_g||_{L^2(\partial\Omega)}  = 2^{-3} $), and compare them to cases where the error on the boundary is smaller ($||\epsilon_f||_{L^2(\Omega)} = 2^{-3}$, $||\epsilon_h||_{L^2(\partial\Omega)} = ||\epsilon_g||_{L^2(\partial\Omega)}  = 2^{-4}$) as shown in figure~\ref{fig:example} to illustrate the effect of lowering the error on the boundaries for both pure cases. We can now use figure~\ref{fig:example} to illustrate how to choose boundary conditions when both Neumann and Dirichlet BCs are accessible. When the domain is large (e.g., $L>10$) and the error on the boundaries is large ($||\epsilon_h||_{L^2(\partial\Omega)} = ||\epsilon_g||_{L^2(\partial\Omega)}  = 2^{-3} $, solid lines in figure \ref{fig:example}), the Neumann boundary conditions yield about twice the error of the Dirichlet boundary. Thus, choosing Dirichlet boundary conditions is best when the the domain is large. However, when the domain is small (e.g., $L<10$), the Neumann BCs yield smaller error. If Neumann BCs are the only choice, one can either improve the experiments with more accurate boundary conditions (e.g.,  green dashed line, $L<3.8$, comparing with the red solid line), or use a smaller domain (e.g.,  blue solid line, $L<1.3$). However, in practice, the scale of the non-dimensionalized flow field is usually large ($L>1$), thus the best choice is accurate Dirichlet BCs with a small flow domain (purple dashed line).  Even for these very simple cases it is complicated to choose the proper BC settings, thus we suggest that users plot their own figure like figure~\ref{fig:example} to design/optimize their own experiments.  A detailed users guide is beyond the scope of this paper and we leave it as the work of a future study.

In this paper, we limited the discussion of the error propagation from the data to the pressure field (denoting as $||f(\bm{u})|| \rightarrow ||p||$), but the error propagation from the velocity field to the data (denoting as $||\bm{u}|| \rightarrow ||f(\bm{u})||$) was not covered {\color{black} due to the greatly increased difficulty of finding solutions.} 

{\color{black} Mathematically,} to bound $||f(\bm{u})||$ with $|| \bm{u}||$ is not an easy task due to the nonlinear terms in the Navier-Stokes equation (e.g., $\bm{u} \cdot \nabla \bm{u}$) making the 2D solution inherently complicated.    {\color{black} On the other hand, even the linear terms (e.g., $\partial \bm{u} / \partial t $) are not bounded terms without additional assumptions.} The 3D version of the propagation of error from the velocity field to the data is related to the well-posedness of the 3D Navier-Stokes equation, which is a Millennium Prize Problem. {\color{black} Physically, a great variety of errors introduced by experiments make this problem even more complicated. For example, particle slip likely introduces high frequency high amplitude local errors; inaccurate calibration introduces low frequency low amplitude global errors, etc. These different types of error introduce different and complicated error propagation phenomena, which we do not fully understand.  For these reasons} we do not expect to make significant progress in this area, at least in 3D. {\color{black} So far, we can only qualitatively explain why the profile of the error and profile of the velocity field are coupled and together dominate the connections between $|| \epsilon_f ||$ and $|| \epsilon_u||$.}

Instead of a full solution to $||\bm{u}|| \rightarrow ||f(\bm{u})||$, we can attempt to calculate the error propagation from the velocity vector field to the data field. These first steps of calculation can provide qualitative intuition of the error propagation in the whole pressure calculation process: 
\begin{equation}
\label{eq:u2f}
\epsilon_f = f(\bm{u+\epsilon_u}) - f(\bm{u}), 
\end{equation}
where $\bm{\epsilon}_u$ is the error vector in the velocity field. Depending on the dimension and non-dimensional numbers in the Navier-Stokes equations (e.g., Reynolds number $Re$, etc.), equation (\ref{eq:u2f}) can be very long, however, the 2D convection term alone should be enough to illustrate the physics. Assuming that $\epsilon_u$ is sufficiently small, we can neglect the second order terms in the error (e.i., $(\partial \epsilon_u/ \partial x)^2$, and $(\partial \epsilon_v/ \partial y)^2$) to approximate \eqref{eq:u2f} as:
\begin{equation}
\label{eq:Ef}
\epsilon_f \approx - 2 \left( \frac{\partial u}{\partial x}\frac{\partial \epsilon_u}{\partial x} + \frac{\partial v}{\partial x}\frac{\partial \epsilon_u}{\partial y} + \frac{\partial u}{\partial y}\frac{\partial \epsilon_v}{\partial x} + \frac{\partial v}{\partial y}\frac{\partial \epsilon_v}{\partial y} \right), 
\end{equation}
where $u$ and $v$ are the velocity components, and $\epsilon_u$ and $\epsilon_v$ are the velocity error in the $x$ and $y$ direction, respectively. Recalling that $\|\epsilon_f\|_{L^2(\Omega)}$  is the source of the error from the velocity field that appears as data in the pressure field calculation (e.g., inequalitites (\ref{eq:DBCboundGeneral}) and (\ref{eq:NBCBoundGeneral})), $||\epsilon_f||_{L^2(\Omega)}$ is calculated by integrating $\epsilon_f^2$ over the whole domain. Utilizing the Cauchy-Schwarz inequality and applying index notation to (\ref{eq:Ef}) we arrive at 
\begin{equation}
\|\epsilon_f\|_{L^2(\Omega)} \leq 2 \left\|\frac{\partial u_i}{\partial x_j}\right\|_{L^2(\Omega)} \left\|\frac{\partial \epsilon_j}{\partial x_i}\right\|_{L^2(\Omega)}.
\label{eq:Efnrom}
\end{equation}
Noticing that the first term, $\left\|{\partial u_i}/{\partial x_j}\right\|_{L^2(\Omega)}$, in \eqref{eq:Efnrom} is actually the gradient of the velocity field, we are be able to obtain some qualitative sense of the reason why the type of the flow affects the error propagation. Physically this means that the velocity gradient directly influences the error level, so for spatially accelerating flow fields the error will inherently be larger. This preliminary discussion is supported by experimental results and physical intuition outlined by \citet{charonko2010assessment}. {\color{black} Similarly, for the error on the boundary, the error in data for the 2D case can be bounded as
\begin{equation}
\label{eq:Egnrom}
\|\epsilon_{g}\|_{L^2(\partial \Omega),i} \leq \epsilon_j \left\|\frac{\partial u_i}{\partial x_j}\right\|_{L^2(\partial \Omega)}  + u_j\left\|\frac{\partial \epsilon_i}{\partial x_j}\right\|_{L^2(\partial\Omega)},
\end{equation}
where $\|\epsilon_{g}\|_{L^2(\partial \Omega),i}$ is the component $i$ ($i=1,2$) of the error on the boundary of the data field. Inequality~(\ref{eq:Egnrom}) shows that the error of the data on the boundary is related to the velocity and velocity gradient, as well as the error and the gradient of the error, which is even more complicated than the case for the error inside the domain (\ref{eq:Efnrom}). We will leave this issue for future studies.}    

We have intentionally made the results of this work unrelated to any specific numerical scheme.  The error bound derived here may be saturated by the worst case scenario with the best numerical implementation. This means that if we solve the pressure equation perfectly with an exact numerical scheme, given a certain level of error in the velocity field, the error in the pressure field will be below the error bound. On the other hand, the numerical error is not considered here, and one may expect larger error than the bounds if the numerical solver is not implemented properly.

One more note on the the non-dimensionalization of the problem may help with practical implementation.
The characteristic length scale of the flow field is exactly the characteristic length ($L^*$) of the Reynolds number, $Re=\rho \bm{u} L^* / \mu$, where $\rho$ and $\mu$ are the  density and dynamic viscosity of the fluid. The pressure can also be related to the characteristic length scale through the non-dimensionalization. For instance, the pressure or error in the pressure can be non-dimensionalized by either a dynamic pressure ($P^*=1 / \rho U^{*2}$, useful for large $Re$ flows), or by a length scale and viscous stresses ($P^*=L^*/\mu U^*$, useful for viscous flows), where $U^*$ is the characteristic velocity of the flow. Thus, the predicted absolute error in the pressure field with real units should be $E_p = || \epsilon_p||_{L^2(\Omega)} P^*$. If we define relative error as $E_p/P^* \times 100\%$, we will see that $|| \epsilon_p||_{L^2(\Omega)}$ actually has physical meaning as a measurement of the relative error of the pressure field. Finally, it isn't necessary to work with the non-dimensionalized Navier-Stokes equation and the pressure Poisson equation as we did here, rather one could re-derive these error bounds dimensionally, but the conclusions would remain the same yet be more difficult to interpret. 

At last, under the framework proposed in this paper, we try to connect two popular categories of methods for PIV-based pressure field calculation: i) pressure Poisson equation based methods, which work with Laplacian of the pressure field derived by applying divergence on a rearranged Navier-Stokes equation (e.g., \citet{de2012instantaneous}); and ii) Navier-Stokes equation based methods, which directly integrate the pressure gradient in the Navier-Stokes equation (e.g., \citet{dabiri2013algorithm} and \citet{liu2006instantaneous}). One may notice that the derivation from the incompressible Navier-Stokes equation to (\ref{eq:NSeq}) according to the statement of the problem is not based on any additional assumptions. This implies that the analysis and the results of the pressure Poisson equation (\ref{eq:NSeq}) in this paper holds for the Navier-Stokes equation based methods too. For example, a large domain accumulates more error in the pressure field from inaccurate velocity measurement, and Dirichlet BCs tends to yield less error than Neumann BCs, etc. The rigorous validation of this point is beyond the scope of this paper, and we sincerely welcome discussion and collaboration on this topic in future studies.

\section{Conclusions}


In this paper, we have analyzed the error propagation dynamics inherent in the calculation of the pressure Poisson equation from velocity data common to many PIV experiments. We emphasize that this work sets up a framework for analyzing the power/level of the error in the pressure field.  The framework is based on a natural idea that the error in the pressure estimation is a combination of the true value and the error; and the measure of these error can be well defined with their $L^2~\&~L^\infty$ norm. Under this framework, we directly analyze the error in the data as non-negligible perturbations to the pressure Poison equation, and have been able to unravel the dynamics that affect error propagation, namely: the shape, area/volume, and boundary conditions of the flow domain, {\color{black} as well as error level in the field and on the boundary. These factors are coupled and make the error propagation dynamics intrinsically complicated even for the fairly simple domains we used in the examples. Based on the error bounds we derived (Inequalities (\ref{eq:DBCboundGeneral}) and (\ref{eq:NBCBoundGeneral})), particular comments can be made for each specific application. However, there are several general conclusions that we would like to emphasize again here: i) The error propagation is dominated by the error inside the data domain when the domain is large, while the error in the pressure calculation is impacted more by the error on the data boundary when the domain is small. ii) The type of boundary conditions significantly and fundamentally affect the error propagation. iii) Particularly, domains with pure Neumann boundary conditions need to satisfy the compatibility condition, which can prove difficult in PIV and may restrict its application. }

This work lays out guidelines for designing experiments (velocity field measurements) that can be used to calculate the pressure field via the pressure Poisson equation.  In engineering practice, the techniques presented can be used to develop \emph{a priori} error estimations of the pressure field to inform the practical side of experiments and minimize the error propagation inherent in calculating pressure fields from velocity fields.




\section{Acknowledgment}
We thank Dr. John Dallon and Dr. Shue-Sum Chow for constructive discussion at the first stage of the research, and Dr. Xiaofeng Liu, and Dr. Roeland De Kat for later advice and discussions. A portion of the research reported in this publication was supported by the National Institute on Deafness and Other Communication Disorders of the National Institutes of Health under award number 5R01DC009616.

\appendix 
\section{Inequalities and notation}
\label{appA}
Before proceeding we note that in addition to the definition of the $L^2$ norm, we have also made use of the $L^\infty$ or sup norm, defined by:
\begin{equation}
\|f\|_{L^\infty(\Omega)} = \sup_{x\in\Omega}|f(x)|.
\end{equation}

We only require three inequalities for the results obtained here, i.e. for bounds on the $L^2$ norm of the error.  Similar calculations can be performed to obtain bounds on the $L^\infty$ norm of the error, but the analysis is far more complicated and hence is omitted.  We also point out that these inequalities are valid only when both sides are finite, i.e. the relevant functions live in the appropriate function spaces.  For further details on such inequalities, we refer to standard textbooks such as \cite{FoiasManleyRosaTemam2001}.
\begin{enumerate}
\item Cauchy-Schwarz:
\begin{equation}
\left|\int_\Omega f(x)g(x)dS\right| \leq |\Omega|\|f\|_{L^2(\Omega)}\|g\|_{L^2(\Omega)}.
\end{equation}
\item Poincare:
\begin{equation}
\|f\|_{L^2(\Omega)} \leq C\|\nabla f\|_{L^2(\Omega)},
\end{equation}
where $C$ is the Poincare constant that depends both on the boundary conditions and the geometry of the domain.  $C$ can also be thought of as the square of the reciprocal of the smallest eigenvalue of the Laplace operator acting on the domain $\Omega$ with the same boundary conditions as those prescribed to $f$.
\item Minkowski (triangle inequality):
\begin{equation}
\|f + g\|_{L^2(\Omega)} \leq \|f\|_{L^2(\Omega)}+\|g\|_{L^2(\Omega)}.
\end{equation}
\end{enumerate}

\section{Error bounds in $L^2$ space}
\label{appB}
\subsection{Dirichlet case} 
\label{sec:DBC_L2_bound} 
Using the principle of superposition and the linearity of the Poisson pressure equation, we can rewrite the solution to equation (\ref{eq:PDE_error}) and (\ref{eq:DBC_error}) as $\epsilon_p = \epsilon_{p(\mathcal{L})} + \epsilon_{p(\mathcal{P})}$, where
\begin{equation}
\begin{split}
& \nabla^2 \epsilon_{p(\mathcal{L})}=0     \quad in \quad \Omega\\
& \epsilon_{p(\mathcal{L})} = \epsilon_h \quad  on  \quad \partial\Omega, \\
\end{split}
\label{eq:PDE_DBC_L}
\end{equation}
and
\begin{equation}
\label{eq:PDE_DBC_P}
\begin{split}
& \nabla^2 \epsilon_{p(\mathcal{P})} =\epsilon_f   \quad in \quad \Omega\\
& \epsilon_{p(\mathcal{P})} = 0 \quad on \quad \partial\Omega. \\ 
\end{split}
\end{equation}

Equation (\ref{eq:PDE_DBC_L}), which is harmonic, satisfies the maximum principle: 
\begin{equation}
\label{eq:maxpcpl}
||\epsilon_{p(\mathcal{L})}||_{L^2(\Omega)} \leq   \sqrt{ \frac{\int_\Omega \|\epsilon_h\|_{L^\infty(\partial\Omega)}^2 dS}{|\Omega|} } = ||\epsilon_{h}||_{L^{\infty}(\partial\Omega)},
\end{equation}
where $|\Omega|$ refers to the area or volume of the region $\Omega$.

Now multiplying \eqref{eq:PDE_DBC_P} by $\epsilon_{p(\mathcal{P})}$ and integrating over the entire domain, we have
\begin{equation}
\label{eq:int}
\int_{\Omega}\epsilon_{p(\mathcal{P})} \nabla^2 \epsilon_{p(\mathcal{P})} \mathrm{d}S=\int_{\Omega}\epsilon_{p(\mathcal{P})} \epsilon_f \mathrm{d}S. 
\end{equation}
Integrating by parts equation \ref{eq:int} yields
\begin{equation}
\label{eq:green}
\oint_{\partial\Omega}\epsilon_{p(\mathcal{P})} \nabla\epsilon_{p(\mathcal{P})} \cdot \mathbf{n} \mathrm{d}L-\int_{\Omega}\nabla\epsilon_{p(\mathcal{P})}^2\mathrm{d}S=\int_{\Omega} \epsilon_{p(\mathcal{P})} \epsilon_f \mathrm{d}S. 
\end{equation}
Substituting homogeneous BCs to equation \eqref{eq:green}, we have
\begin{equation}
\label{eq:green1}
\int_{\Omega}\nabla\epsilon_{p(\mathcal{P})} ^2\mathrm{d}S=-\int_{\Omega}\epsilon_{p(\mathcal{P})}  \epsilon_f \mathrm{d}S.
\end{equation}
This can be rewritten as 
\begin{equation}
\label{eq:H1}
\|\nabla\epsilon_{p(\mathcal{P})}\|_{L^2(\Omega)}^2= -\frac{1}{|\Omega|} \int_{\Omega}\epsilon_{p(\mathcal{P})} \epsilon_f \mathrm{d}S.
\end{equation}
Applying Poincare and Cauchy--Schwarz inequalities \eqref{eq:H1} yields 
\begin{equation}
\label{eq:DBCinfield}
||\epsilon_{p(\mathcal{P})}||_{L^2(\Omega)}^2 \leq C_D ||\epsilon_{p(\mathcal{P})}||_{L^2(\Omega)} ||\epsilon_f||_{L^2(\Omega)}, 
\end{equation}
where, $C_D$ is the Poincare constant for the Dirichlet boundary value problem.

Combining \eqref{eq:maxpcpl} and \eqref{eq:DBCinfield}, and using the Minkowski inequality we have 
\begin{equation}
\label{eq:DBC_fld_bdry}
||{\epsilon_p}||_{L^2(\Omega)} =||{\epsilon_{p(\mathcal{P})}+\epsilon_{p(\mathcal{L})}}||_{L^2(\Omega)} \leq C_D ||{\epsilon_f}||_{L^2(\Omega)}+ ||{\epsilon_h}||_{L^{\infty}(\partial\Omega)}. 
\end{equation}

\subsection{Neumann BCs}
\label{NBC_L2}
Similar to the Dirichlet case, the Poisson equation with non-homogeneous Neumann BCs \ref{eq:PDE_error} and \ref{eq:NBC_error} can be solved by superimposing a Poisson equation with homogeneous BCs
\begin{equation}
\begin{split}
& \nabla^2 \epsilon_{p(\mathcal{P})} = \epsilon_f \quad in \quad \Omega\\
& \nabla \epsilon_{p(\mathcal{P})} \cdot \mathbf{n} = 0 \quad on \quad \partial\Omega , \\ 
\end{split}
\label{eq:PDE2-1}
\end{equation}
and a Laplace equation with non-homogeneous BCs
\begin{equation}
\begin{split}
& \nabla^2 \epsilon_{p(\mathcal{L})} = 0    \quad in \quad \Omega\\
& \nabla \epsilon_{p(\mathcal{L})} = \epsilon_g \quad on \quad \partial\Omega. \\
\end{split}
\label{eq:PDE2-2}
\end{equation}

Solutions of \eqref{eq:PDE2-1} exist only when the compatibility condition
\begin{equation}
\int_{\Omega} \epsilon_f dS = 0
\end{equation}
is satisfied, which means the mean value of the error in the data is assumed zero. 
With this in mind, we multiply \ref{eq:PDE2-1} with $\epsilon_{p(\mathcal{P})}$ and integrate over the entire domain, integrating by parts and using the homogeneous boundary conditions to arrive at:
\begin{equation}
\label{eq:green2}
||\nabla\epsilon_{p(\mathcal{P})}||_{L^2(\Omega)}^2 = |\Omega|^{-1}\int_{\Omega}|\nabla\epsilon_{p(\mathcal{P})}|^2\mathrm{d}S = - |\Omega|^{-1} \int_{\Omega}\epsilon_{p(\mathcal{P})} \epsilon_f \mathrm{d}S .
\end{equation}

Applying the Cauchy--Schwarz and Poincare inequalities, we see that 
\begin{equation}
\label{eq:CS1}
\dfrac{1}{C_N}||{\epsilon_{p(\mathcal{P})}-\bar{\epsilon}_{p(\mathcal{P})}}||_{L^2(\Omega)}^2 \leq ||{\nabla \epsilon_{p(\mathcal{P})}}||_{L^2(\Omega)}^2 \leq ||{\epsilon_{p(\mathcal{P})}}||_{L^2(\Omega)} ||{\epsilon_f}||_{L^2(\Omega)},
\end{equation}
where $C_N$ is the Poincare constant for these boundary conditions and $\bar{\epsilon}_{p(\mathcal{P})}=\int_{\Omega}\epsilon_{p(\mathcal{P})} \mathrm{d}S$, the mean of the pressure field. The compatibility condition on the boundary condition allows us to assume that $\bar{\epsilon}_{p(\mathcal{P})}$ vanishes, and thus $||{\epsilon_{p(\mathcal{P})}}||_{L^2(\Omega)}$ can be bounded as
\begin{equation}
\label{eq:bound_N}
||\epsilon_{p(\mathcal{P})}||_{L^2(\Omega)} \leq C_N ||{\epsilon_f}||_{L^2(\Omega)}. 
\end{equation}

A similar approach to \eqref{eq:PDE2-2} yields 
\begin{equation}
\label{eq:green3}
\|{\nabla\epsilon_{p(\mathcal{L})}}\|_{L^2(\Omega)}^2 = |\Omega|^{-1}\int_{\Omega}|\nabla \epsilon_{p(\mathcal{L})}|^2 \mathrm{d}S = |\Omega|^{-1}  \oint_{\partial \Omega}\epsilon_{p(\mathcal{L})} \epsilon_g \mathrm{d}L.
\end{equation}
Using the Poincare inequality twice on the domain and boundary, respectively, 
\begin{equation}
\label{eq:long2}
\begin{aligned}
\dfrac{1}{C_N}||{\epsilon_{p(\mathcal{L})}-\bar{\epsilon}_{p(\mathcal{L})}}||_{L^2(\Omega)}^2 \leq ||{\nabla\epsilon_{p(\mathcal{L})}}||_{L^2(\Omega)}^2 
& \leq \frac{|\partial\Omega|}{|\Omega|} ||{\epsilon_{p(\mathcal{L})}}||_{L^2(\partial\Omega)} ||{\epsilon_g}||_{L^2(\partial\Omega)}\newline \\
& \leq C_{NB} \frac{|\partial\Omega|}{|\Omega|} ||{ \nabla \epsilon_{p(\mathcal{L})}}||_{L^2(\partial\Omega)}  ||{\epsilon_g}||_{L^2(\partial\Omega)}\\
&= C_{NB} \frac{|\partial\Omega|}{|\Omega|} ||{\epsilon_g}||^2_{L^2(\partial\Omega)}.
\end{aligned}
\end{equation}
Assuming $||\bar{\epsilon}_{p(\mathcal{L})}||_{L^2(\Omega)}$ vanishes, and combining \eqref{eq:bound_N} and \eqref{eq:long2} we have 
\begin{equation}
\label{eq:NBC_fld_bdry}
||{\epsilon}_p||_{L^2(\Omega)} = ||\epsilon_{p(\mathcal{P})}+\epsilon_{p(\mathcal{L})}||_{L^2(\Omega)}\leq C_N ||{\epsilon_f}||_{L^2(\Omega)} + \sqrt{C_N C_{NB} } ||{\epsilon_g}||_{L^2(\partial\Omega)} \sqrt{\frac{|\partial\Omega|}{|\Omega|}}, 
\end{equation}
where, $C_{NB}$ is the Poincare constant for the specified boundary conditions.  We note that the constant on the boundary is due to the compatibility condition, i.e. the error on the boundary is mean zero.

\section{Calculation of the Poincare constants}
\label{appC}
For a smooth and bounded domain, the optimal (minimum) Poincare constant for the Laplace operators is the reciprocal of the first eigenvalue of the BVP problem. As examples, we list the exact Poincare constant for the simple cases illustrated in the paper. For the pure Dirichlet BC case in $M \times N$ domain, the first eigenvalue is $\lambda_1 = \pi^2/M^2 + \pi^2/N^2$, and thus the Poincare constant is $C_D = \lambda_1^{-1} = \pi^2(MN)^2/(M^2+N^2)$. Similarly, for the pure Neumann boundary case, the optimal Poincare constant is $C_N = \max \left( M^2/\pi^2,N^2/\pi^2 \right)$, and for the boundary, $C_{NB} = 2\left( M+N \right)/\pi$. The exact optimal Poincare constant calculation is generally difficult for an arbitrary domain, however, Rayleigh quotient and Rayleigh quotient iteration can be employed to numerically estimate the optimal Poincare constant. See \cite{Gould1995} for one approach to estimating the eigenvalues of such operators.

\bibliographystyle{dcu}
\bibliography{library}
\end{document}